\documentclass[fleqn,twoside]{article}
\usepackage{espcrc2}
\usepackage{graphicx}

\usepackage{graphicx}
\usepackage[figuresright]{rotating}

\newcommand{\AmS}{{\protect\the\textfont2
  A\kern-.1667em\lower.5ex\hbox{M}\kern-.125emS}}

\newcommand \VEV [1] {\left\langle{#1}\right\rangle}

\newcommand{\beq}{\begin{eqnarray}}
\newcommand{\eeq}{\end{eqnarray}}

\newcommand{\LMS}{\Lambda_{\overline{\rm MS}}}

\newcommand{\gev}{\rm GeV}

\def\npb#1#2#3{Nucl.\ Phys.\ {\bf B#1} (#2) #3}

\title{An Instanton Picture O.P.E. $\VEV{A^2}$ Condensate?}
{\small
\author{Ph. Boucaud\address[LPT]{Laboratoire de Physique Th\'eorique, 
	Universit\'e de Paris XI, Batiment 210, 91405 Orsay-Cedex, France.}, 
	F. De Soto\address[US]{Dpto. de F\'{\i}sica At\'omica Molecular y 
	Nuclear, Universidad de Sevilla. Avda. Reina Mercedes s/n, 41012 
	Sevilla, Spain.}, A. Donini\address[INFN]{I.N.F.N., Roma I and Dip. 
	Fisica, Universit\'a di Roma "La Sapienza", P. le A. Moro 2, 00185 
	Rome, Italy}, J.P. Leroy\addressmark[LPT], A. Le Yaouanc\addressmark[LPT], 
	J. Micheli\addressmark[LPT], H. Moutarde\address[EP]{Centre de Physique 
	Th\`eorique Ecole Polytechnique, 91128 Palaiseau Cedex, France}, 
	O. P\`ene\addressmark[LPT], J. Rodr\'{\i}guez-Quintero\address[UHU]{Dpto. 
	de F.A., E.P.S. La R\'abida, Universidad de Huelva, 21819 Palos de 
	la Fra., Spain.}}
}

\begin{document}

\begin{abstract}
Gluon two- and three-point Green Functions computed in Landau gauge
from the  lattice show the existence of power corrections to the
purely perturbative expressions, that can be explained through an
Operator Product Expansion as the influence of a non gauge invariant
mass  dimension two condensate. The relationship of this condensate
with topological properties of QCD, namely instantons, will be
studied, giving a first estimate of the contribution of instantons
to this condensate based in the direct lattice  measure, after a
cooling process, of the instanton liquid properties.
\vspace{1pc}
\end{abstract}

\maketitle

\section{QCD coupling constant, O.P.E. and $\VEV{A^2}$ condensate.}

Lattice calculations of the QCD coupling constant and gluon propagator 
based in the Green Functions Method \cite{alles}, suggest the necessity 
to add power corrections to the purely perturbative 
expressions to  correctly describe their running \cite{power}. An 
Operator Product Expansion (O.P.E.) analysis of the Green functions in
Landau gauge\footnote{In the lattice we will work in the minimum $A^2$ 
Landau gauge, $\partial_\mu A_\mu =0$, so all gauge dependent quantities, 
will be expressed in this particular gauge.} relates this power 
corrections to the existence of a non-perturbative $\VEV{A^2}$ 
condensate \cite{ope}, through expressions:

\beq
G^{(2)}_{O.P.E.}(p^2)=G^{(2)}_{Pert.}(p^2)+c\frac{\VEV{A^2}_{R,\mu}}{p^2},\nonumber \\
\alpha_s^{O.P.E}(p^2)=\alpha_s^{Pert.}(p^2)+c'\frac{\VEV{A^2}_{R,\mu}}{p^2},
\label{ope}
\eeq

\noindent where perturbative expressions are developed at three loops, and the
functions $c$ and $c'$ include the Wilson coefficient of the expansion, and 
the anomalous dimension of the  condensate at leading logarithm.

By performing a combined fit of lattice results to expressions in (\ref{ope}), 
in two different ${\rm MOM}$ schemes, a value of $\LMS$ is extracted, in 
fairly good agreement with the one obtained by the ALPHA collaboration 
\cite{alpha}, by a completely  different method. A value of the condensate 
comes out from the analysis. The physical meaning of this condensate is  still
an open question, and a  lot of work is being devoted to its study during last
years, for example,  in relation to confinement \cite{Kondo}. The aim in
this work will be to study the possible semiclassical contribution to this
condensate coming from  instantons, and whether they might explain the presence
of power corrections in  Green Functions.

\section{The role of instantons.}

Instantons have been extensively studied as a possible description of the QCD vacuum 
(See \cite{shuryak} for a general overview), and so 
as a major source of QCD properties at low energies. In relation with the 
aim of this work, an ensemble of non-interacting instantons (${\rm I}$) and 
antiinstantons (${\overline{\rm I}}$) in Landau gauge would give a 
contribution to the  $\VEV{A^2}$ condensate;

\beq
\VEV{A^2}_{\rm inst} \approx 
\frac{N}{V} \int d^4x A_\mu^a(x) A_\mu^a(x) = 12 \pi^2 \rho^2 n ,
\label{a2i}
\eeq

\noindent where $A_\mu^a(x)$ is the standard 't Hooft Polyakov instanton 
gauge field \cite{thooft}, $\rho$ the average radius, and 
$n=\frac{N_I+N_{\overline{\rm I}}}{V}$ the density. 

If we accept the phenomenological values assigned to $n$ and $\rho$ by the 
Instanton Liquid Model \cite{shuryak} ($n\sim 0.5 fm^{-4}$ and 
$\rho \sim 1/3 fm$), the instantonic contribution will be 
$\VEV{A^2}_{Inst.}\sim 0.5 \gev^2$. We will perform, however,
our own analysis, thus testing the latter approach.

\subsection{Cooling.}

In principle, a direct measure of $A^2$ in the lattice should be possible, but 
the presence  of the UV divergent part is hardly separable from the soft, 
instantonic one. The other possibility is to perform a cooling procedure, 
that will  allow us to compute the number and size of instantons, giving 
an indirect  measure of the $A^2$ through (\ref{a2i}).

We will use the traditional cooling method \cite{teper}, even if it introduces a number 
of known biases, as ${\rm I}-{\overline{\rm I}}$ annihilation, and a modification of
instanton sizes and lattice spacing. The approach proposed here is to compute 
instanton properties for different number of cooling sweeps, and extrapolate back
to the thermalised situation in order to recover their physical meaning\footnote{
The use of improved cooling methods, as the one developed in \cite{garciaperez}, could 
improve this approach, as radii evolution is minimised, but ${\rm I}-{\overline{\rm I}}$ 
annihilation is unavoidable, so the extrapolation will be anyway necessary.}.

\subsection{Shape Recognition.}

Instantons will be localised in cooled lattices via a geometrical method (Described 
in \cite{instanton}.) that accepts a topological charge lump as an instanton when the 
ratio of the integral over a given fraction of the topological charge at the maximum, 
$\alpha$, and its theoretical counterpart, $\epsilon$, is $\sim 1$, for a range of 
values of $\alpha$:

\beq
\epsilon=\frac{\int_{x/\frac{|Q_\rho(x)|}{|Q_\rho(0)|}\ge \alpha} d^4x Q_\rho(x)}
{1-3\alpha^{1/2}+2\alpha^{3/4}}
\eeq

Once the lump has been identified as an instanton, the radius will be computed 
from the size of the cluster where the integral has been developed.

\subsection{A naive model of annihilation.}

With the method outlined above, we compute the density and size of instantons in a lattice, 
for different numbers of cooling sweeps, $n_c$, obtaining values with a strong dependence on 
$n_c$ (See figure), that avoids to obtain any physical information at fixed $n_c$.

As a first approach to the understanding of this evolution, we will make a 
simple model, where instantons annihilate with antiinstantons (Being so 
$\Delta N=N_I-N_{\overline{I}}$ a constant) proportionally to their packing ratio, 
and to the number of antiinstantons, 
so that the equation for the evolution of $N=N_I+N_{\overline{I}}$ is:

\beq
\frac{\partial N}{\partial n_c}\ = - \frac{\lambda}{2V}\rho^4(n_c) (N(n_c)^2-\Delta N^2).
\label{nnc}
\eeq

\noindent If we assume $\rho(n_c)=cte$, the solution of Eq. (\ref{nnc}) will give 
$N(n_c)\sim\frac{N(0)}{1+\lambda n_c}$, the expression used in \cite{instanton}, 
as a first order approach, but our cooling procedure modifies instanton's size 
(See figure), in a way thah we phenomenologically parametrise as:

\beq
\rho(n_c)\ =\ \rho(0) (1 + a \ln( 1 + n_c ) )\ .
\label{rnc}
\eeq

\noindent We will include (\ref{rnc}) in equation (\ref{nnc}), with $\rho(0)$ the extrapolated 
radius at the thermalised situation and $a$ a constant to determine.

After performing a combined fit of our lattice results to the expressions 
(\ref{rnc}) and the one coming from the integration of (\ref{nnc}), we can fix
the initial values of the density, $n(0)$ and the radius $\rho(0)$, and the 
two constants that govern the evolution, $\lambda$ and $a$.

\begin{figure}[ht]
\begin{center}
\begin{tabular}{c}
\includegraphics[width=15pc]{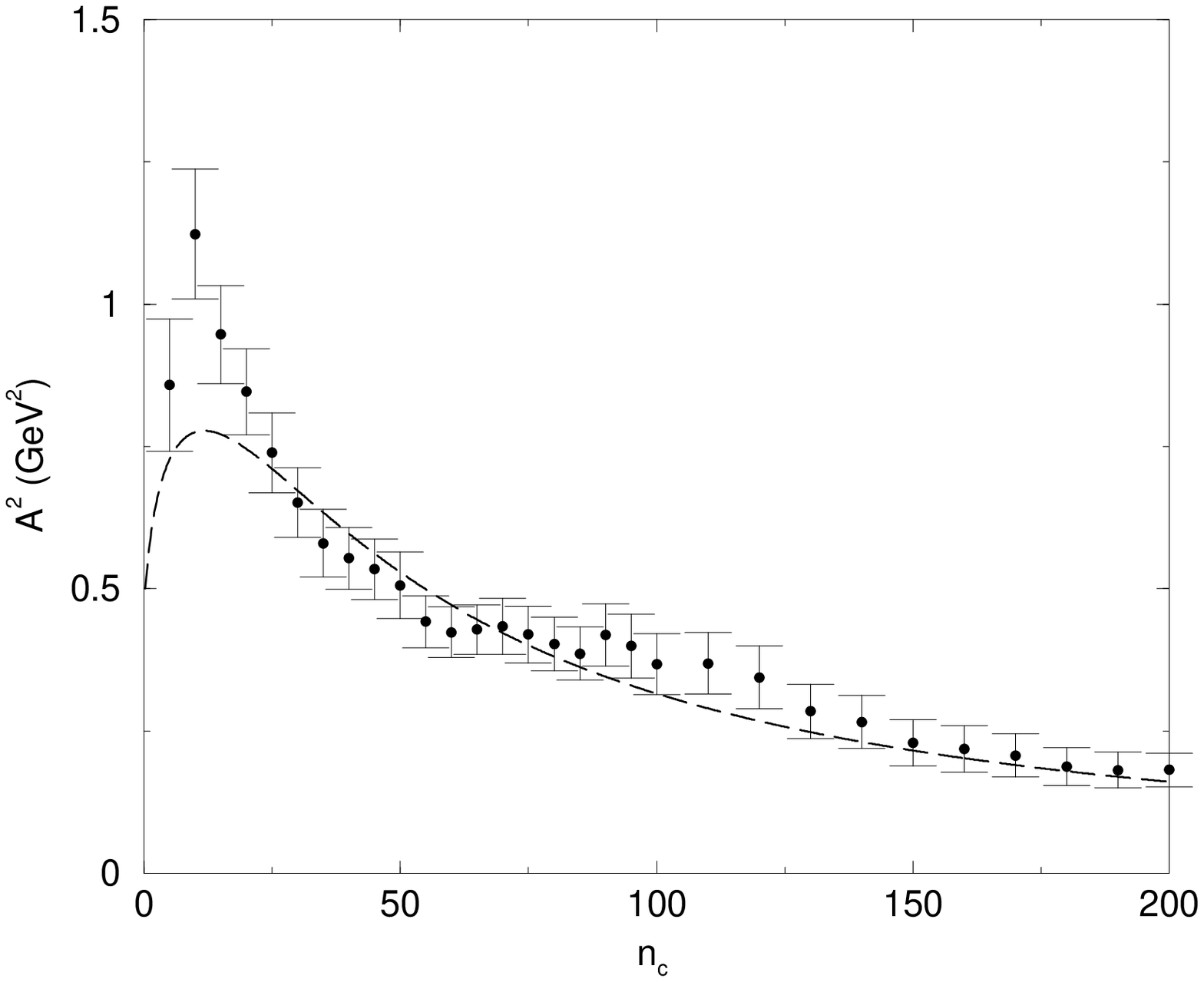}  \\
\includegraphics[width=15pc]{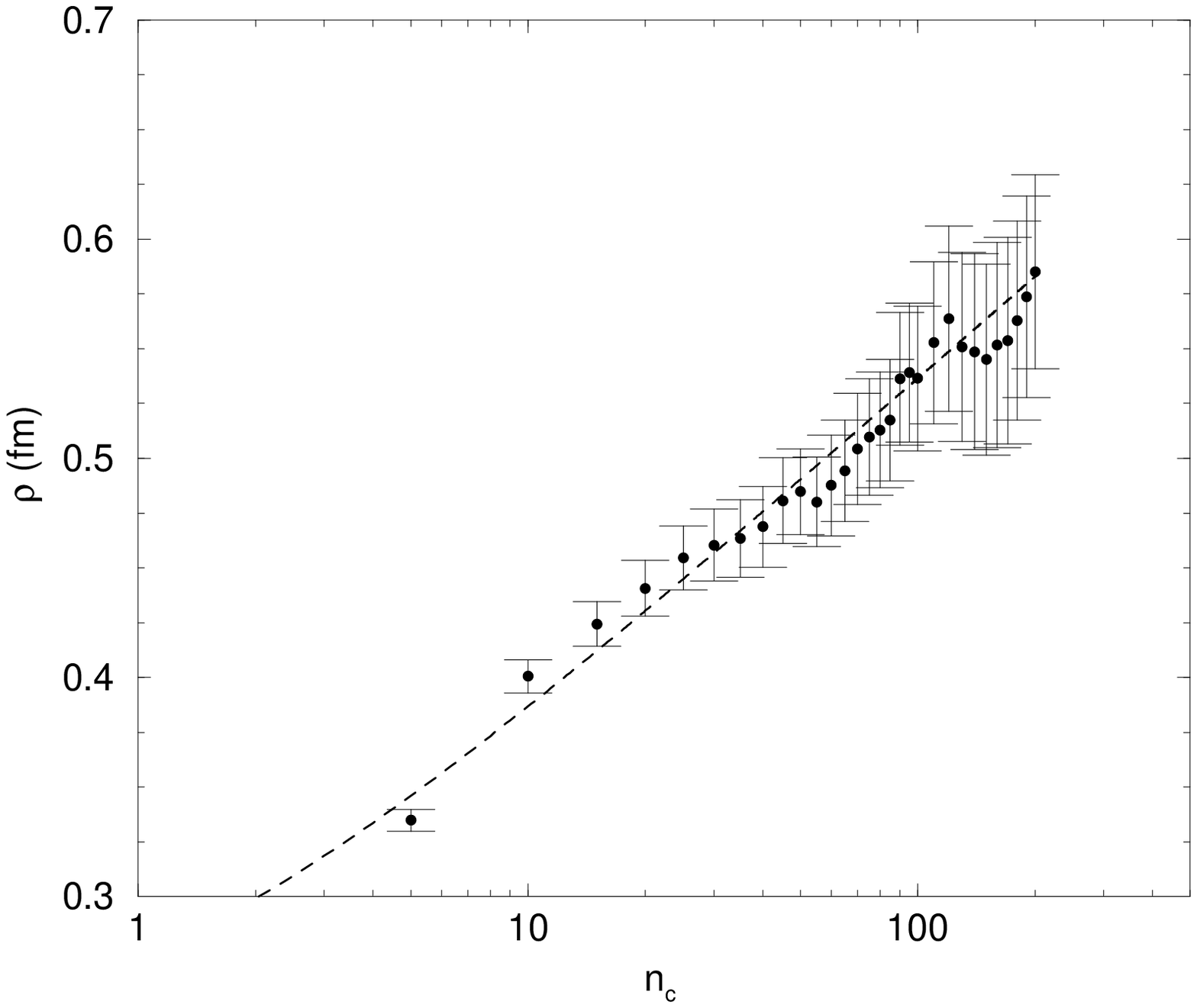} \\
\vspace*{-1cm}
\end{tabular}
\caption{\small {\it Results of the combined fit for the instanton density and radius 
as a function of the number of cooling sweeps for a $24^4$ lattice at $\beta$=6.0.}}
\label{Fign}
\end{center}
\end{figure} 

\vspace*{-1cm}

\section{Results and conclussion.}

The result of the combined fit gives a value of the instantonic 
contribution to $\VEV{A^2_{Ins}}\sim 0.4 \gev^2$, however
the result of the extrapolation is highly dependent on the value 
of $\rho$, which due to the logarithmic behaviour is hardly reliable.
We therefore prefer the value at the maximum, 1.12(11) $\gev^2$, as a 
crude estimation of $\VEV{A^2_{Ins}}$.

This semiclassical evaluation of $\VEV{A^2}$, which does not run with the scale,
is difficult to relate to that appearing in the O.P.E. expansion, which does 
depend on the renormalisation scheme and scale. The typical scale of instantons 
is $\rho^{-1}\sim 0.7\gev$. Unluckly it is not possible to run the $\VEV{A^2_{O.P.E.}}$
to such a low energy, where pertubative QCD is not valid. The lowest reacheble 
energy scale is $2.6\gev$ \cite{ope,instanton};

\beq
\VEV{A^2_{O.P.E.}(2.6\gev)}=1.4(3)(3)\gev^2 ,
\eeq

\noindent the first error coming from the OPE determination of 
the condensate renormalised at 10 $\gev$, and the second from 
higher orders in the running.

Keeping in mind the level of uncertainty of these calculations, 
we can nevertheless claim a rather encouraging agreement between
the instantonic contribution to the condensate and the one
computed from the running of the Green Functions.

\section*{Acknowledgements.}

F. de Soto wants to thank L.P.T. for its warm hospitality and Fundaci\'on
C\'amara and the MCYT (Contract BFM2002-03315) for financial support.

\end{document}